\documentclass{aa}  

\usepackage{xcolor}
\usepackage{graphicx}
\usepackage{txfonts}
\newcommand\isotope[2]{\textsuperscript{#2}#1}
\newcommand\lvmr[1]{\ensuremath{ \log{ \left(\mathrm{VMR_{#1}}\right) } }}
\newcommand\kp{\ensuremath{K_\mathrm{p}}}
\newcommand\vsys{\ensuremath{v_\mathrm{sys}}}
\newcommand\vpl{\ensuremath{v_\mathrm{p}}}
\newcommand\vbary{\ensuremath{v_\mathrm{bary}}}
\newcommand\sig{\ensuremath{\sigma}}
\newcommand\snr{\ensuremath{\mathrm{S/N}}}
\newcommand\teq{\ensuremath{T_\mathrm{eq}}}

\newcommand{\kms}{\ensuremath{\mathrm{km \ s^{-1}}}}
\newcommand{\vpair}[3]{${\ensuremath{ \left(\kp{},\ \vsys{} \right) #1 \left(#2,\ #3 \right)\ \kms{} } }$}
\newcommand{\plez}{Plez98}
\newcommand{\plezN}{Plez12}
\newcommand{\sysrem}{\textsc{SysRem}}
\newcommand{\toto}{ExoMol \textsc{Toto}}
\newcommand{\iraf}{\textsc{IRAF}}
\newcommand{\ptr}{\textit{petitRADTRANS}}

\begin{document} 

   \title{Is TiO emission present in the ultra-hot Jupiter WASP-33b?\\ A reassessment using the improved \toto{} line list}

   \author{
            Dilovan B. Serindag\inst{1},
            Stevanus K. Nugroho\inst{2},
            Paul Molli\`{e}re\inst{3},
            Ernst J. W. de Mooij\inst{2},
            Neale P. Gibson\inst{4},
            \and
            Ignas A. G. Snellen\inst{1}
          }

   \institute{
                Leiden Observatory, Leiden University, Postbus 9513, 2300 RA Leiden, The Netherlands
                \and
                Astrophysics Research Centre, School of Mathematics and Physics, Queen’s University Belfast, University Road,\\Belfast BT7 1NN, United Kingdom
                \and
                Max-Planck-Institut f\"{u}r Astronomie, K\"{o}nigstuhl 17, 69117 Heidelberg, Germany
                \and
                School of Physics, Trinity College Dublin, The University of Dublin, Dublin 2, Ireland
             }
             
    \titlerunning{A reassessment of TiO emission in WASP-33b using \toto{}}
    \authorrunning{Serindag et al. (2020)}

   \date{}
 
  \abstract
   {
    Efficient absorption of stellar ultraviolet and visible radiation by TiO and VO is predicted to drive temperature inversions in the upper atmospheres of hot Jupiters. However, very few inversions or detections of TiO or VO have been reported, and results are often contradictory.
   }
   {
    Using the improved \toto{} line list, we searched for TiO emission in the dayside spectrum of WASP-33b using the same data in which the molecule was previously detected with an older line list at 4.8\sig{}. We intended to confirm the molecular detection and quantify the signal improvement offered by the \toto{} line list.
   }
   {
    Data from the High Dispersion Spectrograph on the Subaru Telescope was extracted and reduced in an identical manner to the previous study. Stellar and telluric contamination were then removed. High-resolution TiO emission models of WASP-33b were created that spanned a range of molecular abundances using the radiative transfer code \ptr{}, and were subsequently cross-correlated with the data.
   }
   {
    We measure a 4.3\sig{} TiO emission signature using the \toto{} models, corresponding to a WASP-33b orbital velocity semi-amplitude of ${\kp{} = 252.9^{+5.0}_{-5.3}\ \kms{} }$ and a system velocity of ${\vsys{} = -23.0^{+4.7}_{-4.6}\ \kms{} }$. Injection-recovery tests using models based on the new and earlier line lists indicate that if the new models provide a perfect match to the planet spectrum, the significance of the TiO detection should have increased by a factor of $\sim$2.
   }
   {
    Although the TiO signal we find is statistically significant, comparison with previous works makes our result too ambiguous to claim a clear-cut detection. Unexpectedly, the new \toto{} models provide a weaker signal than that found previously, which is offset in \kp{}-\vsys{} space. This sheds some doubt on both detections, especially in light of a recently published TiO non-detection using a different dataset.
   }


   \maketitle
%

\section{Introduction} \label{sec:intro}

    The first exoplanet discovered orbiting a main-sequence star was not only significant as a milestone in the search for such objects. As a gas-giant planet with a mass comparable to that of Jupiter but an orbital period of less than ten days, 51 Peg b represented a new class of planet with no analog in the Solar System \citep{Mayor1995}. The discovery of hundreds of these so-called hot Jupiters with such small orbital separations has naturally raised questions about the effects of intense stellar radiation on planetary atmospheres \citep{Hubeny2003,Fortney2008}. It is predicted that for sufficiently high levels of stellar irradiation, gaseous TiO and VO can persist in the upper atmospheres of hot Jupiters. Efficient absorption of ultraviolet and visible radiation by these molecules would lead to an inversion in the pressure-temperature profile---that is, an increase in temperature with decreasing pressure. The extent and prevalence of such inversions are important to understanding the various radiative, chemical, and dynamical processes at work in the atmospheres of hot Jupiters.
    
    Evidence for thermal inversions and spectral signatures from TiO and VO are scarce in the literature, and results are often ambiguous and sometimes even contradictory. In particular, the relation between thermal inversions and the presence of TiO and VO is not evident. WASP-121b is an illustrative example. \citet{Evans2017} resolve near-infrared H$_2$O emission in secondary eclipse spectra, resulting in an unambiguous detection of a thermal inversion. While the inversion in WASP-121b is supported by TESS phase curve and eclipse photometry \citep{Daylan2019,Bourrier2020}, no spectroscopic evidence for TiO has been found \citep{Evans2018,Merritt2020}. Evidence for VO is present in both eclipse \citep{Evans2017,MikalEvans2019} and transit \citep{Evans2018} spectra at low resolution, but not in high-resolution transmission spectra \citep{Merritt2020}. \citet{Gibson2020} detect Fe {\sc I} in high-resolution transmission spectra and suggest the inversion in WASP-121b may instead be driven by neutral iron.
    
    Similarly, CO emission in low-resolution eclipse observations of WASP-18b indicates the presence of an inversion layer, but no evidence for TiO or VO has been found \citep{Sheppard2017}. \citet{Arcangeli2018} confirm the inversion and suggest that H$^{-}$ opacity may play an important role in driving the inversion, instead of TiO or VO. Conversely, WASP-19b has a possible TiO detection but no evidence for a temperature inversion. \citet{Sedaghati2017} report a strong TiO signal in transmission at low resolution, while \citet{Espinoza2019} do not.  
    
    Currently, WASP-33b is the only exoplanet with a reported temperature inversion and TiO emission detection based on high-dispersion spectral observations. Orbiting its A5 host star in a 1.2-day period, the ultra-hot Jupiter WASP-33b (${\teq{} \sim 2700 \ \mathrm{K}}$) is one of the hottest known exoplanets. \citet{Haynes2015} find a temperature inversion necessary to interpret near-infrared \textit{Hubble} Space Telescope eclipse spectra, attributing excess flux at shorter wavelengths to TiO emission. Using cross-correlation techniques on high-resolution optical spectra taken of the planet's dayside with the Subaru Telescope, \citet{Nugroho2017} detect TiO in emission at the 4.8\sig{} level, assuming the inverted pressure-temperature profile from \citet{Haynes2015}. This is both the first high-resolution TiO or VO detection and the first high-resolution detection of a thermal inversion. However, \citet{Herman2020} analyzed similar quality high-resolution optical transmission and dayside spectra of WASP-33b taken with the Canada–France–Hawaii Telescope (CFHT) and do not detect a TiO signal. More recently, \citet{Nugroho2020} have detected Fe {\sc I} emission from WASP-33b in the same Subaru data and suggest neutral iron may also contribute to driving the inversion.
    
    In this paper, we used the new and more accurate \toto{} line list for TiO from \citet{McKemmish2019} to reanalyze the data of WASP-33b from the High Dispersion Spectrograph on Subaru, used by \citet{Nugroho2017} to find TiO emission in its dayside spectrum. We anticipated this improved line list would enable a significantly stronger TiO detection. In Section \ref{sec:data} we present this data of WASP-33b and discuss our methodology for detecting TiO in Section \ref{sec:methods}. We present and discuss our tentative detection in Sections \ref{sec:results} and \ref{sec:discuss}.
    

\section{Subaru data of WASP-33} \label{sec:data}

    For our analysis, we used archival spectra\footnote{Proposal ID: S15B-090; PI: H. Kawahara} of WASP-33 taken on 26 October 2015 using the High Dispersion Spectrograph (HDS) on the Subaru Telescope, first presented in \citet{Nugroho2017}. A slit width of $0\farcs2$ provided a resolving power ${\mathcal{R} = \lambda/\Delta\lambda = 165\,000}$ (${\Delta v = 1.8 \ \kms{}}$), with a pixel sampling of 0.9 \kms{}. Fifty-two observations with integration time ${600\ \mathrm{s}}$ were taken using both the blue and red HDS CCDs covering 6164--7396 \AA{} and 7685--8810 \AA{}, respectively. The observation midpoints span orbital phases ${\phi=0.207 \hbox{--} 0.539}$, with the final 15 exposures taken with the planet in occultation.
    
    The extraction of the one-dimensional HDS spectra is identical to that described in Sections 2.3 and 2.4 of \citet{Nugroho2017}. Bias correction, bad-pixel masking, non-linearity correction, background subtraction, and flat fielding were performed on the blue and red CCD frames using \iraf{}. Subsequently, 18 blue and 12 red orders were extracted and continuum normalized, and a wavelength solution was determined by comparison with Th-Ar frames. A $\sim$0.1-pixel drift in the wavelength solution was identified by comparing the positions of strong telluric lines over the observing night, and corrected by spline interpolation. A 5\sig{} outlier clipping was then applied to each wavelength bin. A full description of these preliminary reduction steps is given in \citet{Nugroho2017}.
    
    While these previous steps are identical to those in the original analysis, \citet{Nugroho2017} performed all further analysis for each order separately. Instead, we combined all blue and red orders into a single spectrum and entirely removed the overlapping wavelength regions, which show poor agreement. These regions amount to 7.5\% of the wavelength bins. Although this removes some flexibility in treating particular orders differently, it also simplifies the analysis. In addition to flagging the telluric O$_2$ A and B bands at 7600 \AA{} and 6900 \AA{}, respectively, we flagged all wavelength bins with a telluric transmission value $\le$ 0.98. These two steps result in a further flagging of 7.9\% and 4.0\% of the wavelength bins, respectively. After dividing each wavelength bin by its median value over all observations, we performed an additional 5\sig{} clipping on each and set all outliers to the bin's median value. 
    
    To further identify and remove tellurics and systematic noise, we performed singular value decomposition (SVD) to deconstruct our data matrix into its constituent components as described in \citet{deKok2013}. We created ten new data matrices in addition to the original, each with a successively weaker SVD component removed. After interpolating each matrix onto a wavelength grid with pixels of width 0.5 \kms{}, we performed a single high-pass filter with full width 20.5 \kms{} (41 pixels) on each observation. We then divided each wavelength bin by its standard deviation over all observations. These last two steps were performed separately on the matrices of each SVD "iteration."
    

\section{Search for TiO emission} \label{sec:methods}

    \subsection{\toto{} opacities for TiO} \label{sec:opacities}

        Recently, \citet{McKemmish2019} published a new line list for TiO as part of the ExoMol project, which aims to calculate complete, accurate line lists for molecular species of interest to exoplanet spectroscopy \citep{Tennyson2012}. This \toto{} line list includes data for all the main TiO isotopologues\footnote{ \isotope{Ti$^{16}$O}{46}, \isotope{Ti$^{16}$O}{47}, \isotope{Ti$^{16}$O}{48}, \isotope{Ti$^{16}$O}{49}, \isotope{Ti$^{16}$O}{50} }, computed using more accurate experimental energy levels. \citet{McKemmish2019} demonstrate the superior quality of their \toto{} line list for the primary isotopologue \isotope{Ti$^{16}$O}{48} compared to the \citet{Plez2012} line list (hereafter, \plezN{}), itself an improvement on the \citet{Plez1998} line list (hereafter, \plez{}) used by \citet{Nugroho2017} in their study of WASP-33b. Specifically, they note the improved completeness of the \toto{} line list, as well as improved positions and strengths of lines when compared to M-dwarf spectra. \citet{Pavlenko2020} show that the \toto{} line lists for the secondary isotopologues also provide an improvement over the \plezN{} line list. We calculated TiO opacities using the \toto{} line list for each main isotopologue on a  high-resolution grid ($\lambda/\Delta\lambda = 10^6$) using the method presented in Appendix A of \citet{Molliere2015}, which improves efficiency by computing the contributions of line cores and wing continua on a high- and low-resolution grid, respectively.
        
\begin{figure*}
    \resizebox{\hsize}{!}
    { \includegraphics{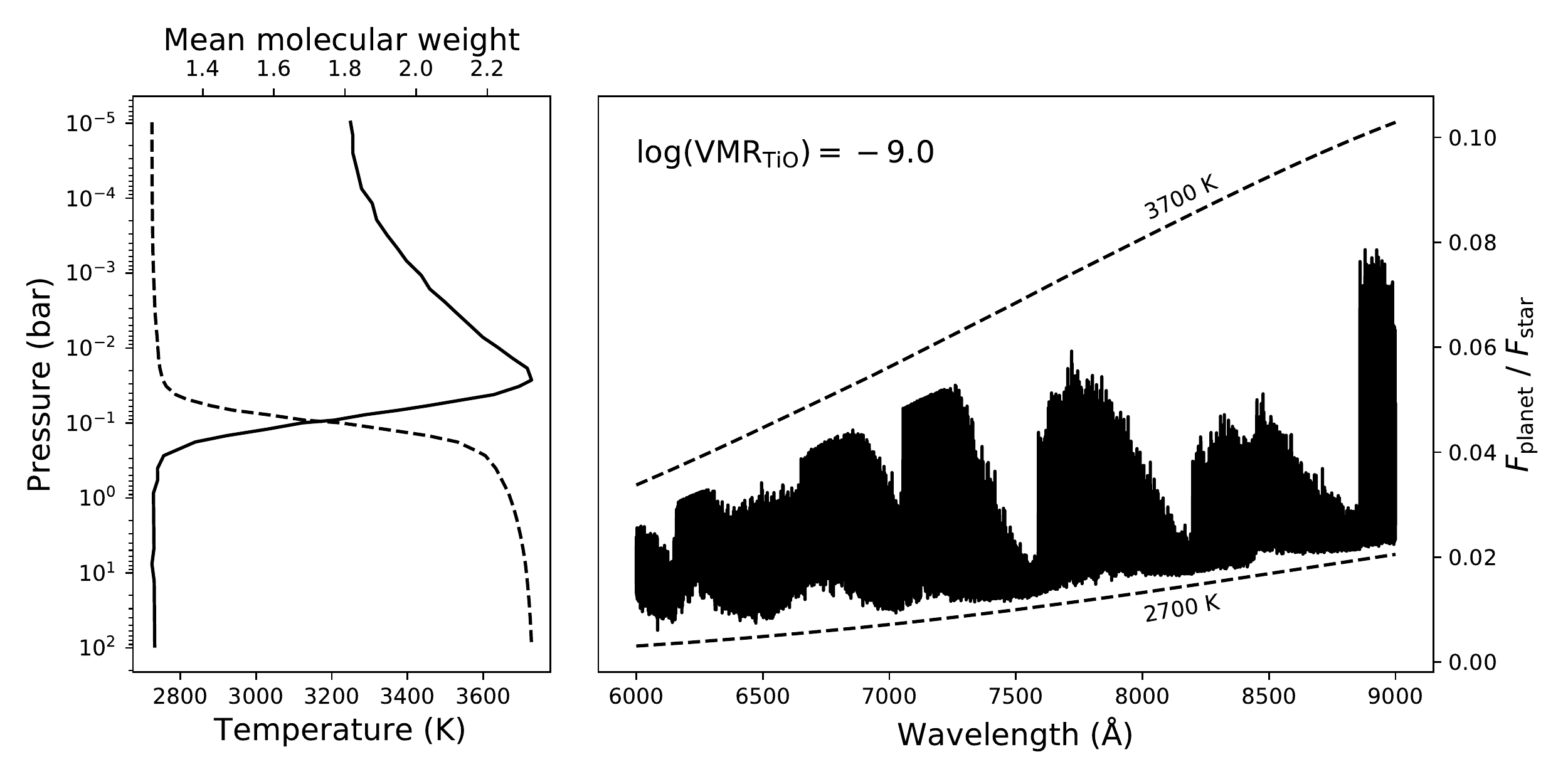} }
    \caption{Left panel: \citet{Haynes2015} pressure-temperature profile with inversion (solid line) and calculated mean molecular weight profile (dashed line) used to model the TiO emission spectrum of WASP-33b. Right panel: Example TiO emission spectrum for WASP-33b assuming a constant ${\lvmr{TiO}=-9.0}$. The planet-to-stellar-continuum flux ratio is plotted (solid line). The dashed lines show the blackbody-to-stellar-continuum flux ratio for the minimum (2700 K) and maximum (3700 K) temperatures of the \citet{Haynes2015} pressure-temperature profile.}
    \label{fig:spectralModel}
\end{figure*}

    \subsection{\ptr{} TiO emission models} \label{sec:petit}
    
        The radiative transfer code \ptr{}\footnote{https://petitradtrans.readthedocs.io} \citep{Molliere2019} was used to model the TiO emission spectrum of WASP-33b. As input for \ptr{}, we adopted the \citet{Haynes2015} pressure-temperature profile with inversion also used by \citet{Nugroho2017}, and calculated a corresponding one-dimensional mean molecular weight profile for 46 atmospheric layers from $10^{2}$ to $10^{-5}$ bar using the equilibrium chemistry code described in Appendix A2 of \citet{Molliere2017} assuming a solar metallicity. Both profiles are plotted in the left panel of Figure \ref{fig:spectralModel}. We included opacity contributions from all five main TiO isotopologues assuming solar abundances. We created a grid of such models, varying the total TiO volume mixing ratio \lvmr{TiO} from -7.0 to -10.0 in steps of 0.2. The ${\lvmr{TiO}=-9.0}$ emission model, scaled by the stellar continuum, is shown as an example in the right panel of Figure \ref{fig:spectralModel}. To maximize any retrieved signal, we broadened these models to the HDS resolving power and performed a high-pass filter to mimic our treatment of the HDS data.
    
    \subsection{Cross-correlation search for TiO emission} \label{sec:ccf}

        To search for TiO emission in WASP-33b, we cross-correlated each model with the noise residuals of each observation over velocities $\pm$600 \kms{} in steps of 0.5 \kms{} relative to the observatory rest frame. The cross-correlation function (CCF) of each observation was normalized such that values of 0 and 1 correspond to no and full match, respectively, and then median subtracted. This process was performed on the data matrix of each SVD iteration separately, resulting in 11 two-dimensional CCF matrices for each model.
    
        To combine any TiO emission signal over time, we shifted the CCF for each observation to the rest frame of the exoplanet, summed the CCFs for each out-of-occultation observation, and performed another median subtraction. This resulted in a one-dimensional CCF for the entire night of observations. The exoplanet velocity relative to the observatory rest frame at orbital phase $\phi$ is given by ${ \vpl(\phi) = \vbary(\phi) + \vsys + \kp\sin{( 2 \pi \phi)} }$, where \vbary{} is the velocity of the observatory relative to the Solar System barycenter, \vsys{} is the velocity of the WASP-33 barycenter relative to that of the Solar System, and \kp{} is the orbital velocity semi-amplitude of WASP-33b. While \vbary{} and the orbital phase for each observation are known, we varied \kp{} and \vsys{} to allow for deviations from those values reported by \citet{Nugroho2017} and to assess the noise properties of the CCF data. We stacked these one-dimensional CCFs into a \kp{}-\vsys{} matrix, where each row is the aligned and summed CCF for a given \kp{} and each column corresponds to a different \vsys{}. We converted the CCF values of each matrix to \snr{} values by dividing each row (constant \kp{}) by its standard deviation. With 11 data matrices (one for each SVD iteration) and 16 model templates, we constructed 176 \kp{}-\vsys{} matrices.
        

\begin{figure}
    \centering
    \includegraphics[width=\hsize]{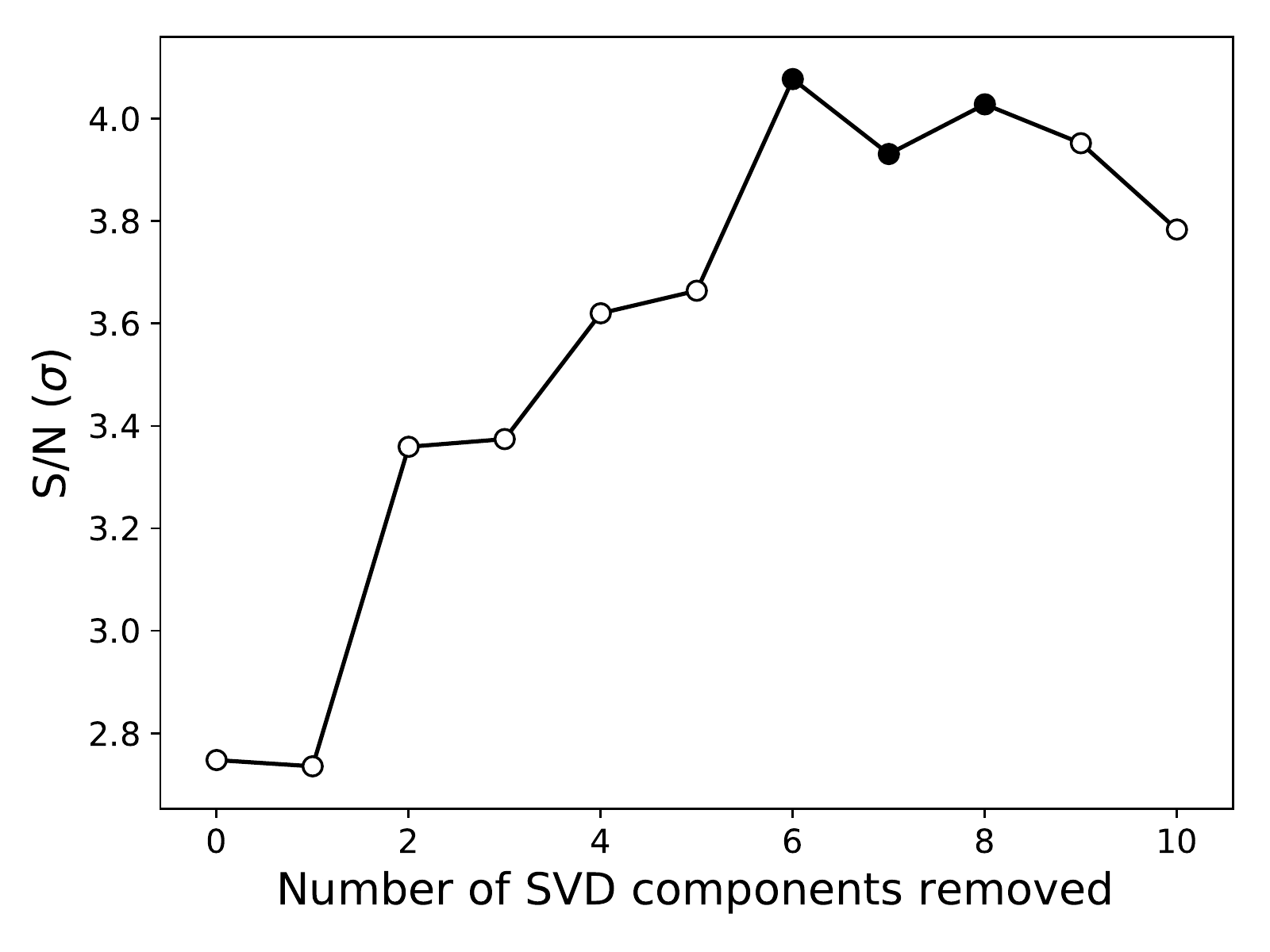}
    \caption{\snr{} of the emission peak around \vpair{\approx}{253}{-23} in the \kp{}-\vsys{} matrices for the ${\lvmr{TiO}=-9.0}$ model and data matrices with various numbers of SVD components removed. The filled circles indicate SVD iterations for which this emission peak is the strongest on the full \kp{}-\vsys{} matrix. Since the \snr{} value exceeds 4\sig{} for SVD iterations 6 and 8, the peak was flagged by our analysis for these iterations.}
    \label{fig:snrPCA}
\end{figure}

\section{Results} \label{sec:results}

    We identified prospective TiO emission by searching for peaks in the \kp{}-\vsys{} matrices that (1) have velocity values within a 3\sig{} box of the \vpair{=}{237.5^{+13.0}_{-5.0}}{-1.5^{+4.0}_{-10.5}} result from \citet{Nugroho2017}, (2) have ${\snr{} \ge 4\sig{} }$, and (3) are the strongest peak on the entire matrix. In the 176 \kp{}-\vsys{} matrices, we found four peaks that satisfy these criteria, all with \vpair{\approx}{253}{-23}: peaks with \snr{} values of 4.05\sig{} and 4.02\sig{} for the SVD6 and SVD8 data matrices and ${\lvmr{TiO} = -8.8}$ model, and peaks with \snr{} values of 4.08\sig{} and 4.03\sig{} for the SVD6 and SVD8 data matrices and ${\lvmr{TiO} = -9.0}$ model.

    Although we could not constrain the TiO volume mixing ratio, based on the slightly stronger \snr{} values, we adopted the ${\lvmr{TiO} = -9.0}$ template as our fiducial model and examined how varying the number of removed SVD components affects these prospective emission peaks. Figures \ref{fig:snrGridPanel1_full_exomol} and \ref{fig:snrGridPanel2_full_exomol} show the full \kp{}-\vsys{} matrices for the fiducial model and the data matrix after each successive SVD iteration. The dashed cyan box in each panel denotes the aforementioned 3\sig{} region within which we searched for potential TiO emission, based on the velocities reported by \citet{Nugroho2017}. The \snr{} and location of the strongest peak on each matrix is noted at the top of each panel and marked by a cyan ring on the matrix. Figures \ref{fig:snrGridPanel1_zoom_exomol} and \ref{fig:snrGridPanel2_zoom_exomol} show the same, but with the \vsys{} axis restricted to $\pm$100 \kms{} for clarity. For SVD iterations 6--8 the strongest peak of each matrix falls within the 3\sig{} box around velocities \vpair{\approx}{253}{-23}. This same emission signal appears present in the \kp{}-\vsys{} matrices for all SVD iterations and becomes relatively prominent after two iterations of SVD.
    
    To better quantify this, we plotted the \snr{} of this emission peak as a function of SVD iteration in Figure \ref{fig:snrPCA}. After two SVD iterations the peak exceeds 3\sig{}, which explains its relative prominence in the \kp{}-\vsys{} matrices for subsequent iterations. The solid markers for SVD iterations 6--8 indicate the peak has the highest \snr{} on the \kp{}-\vsys{} matrix for those iterations. That the strength of the peak falls below 4\sig{} for the SVD7 case explains why it was not initially flagged in our search of all 176 \kp{}-\vsys{} matrices.
    
\begin{figure}
    \centering
    \includegraphics[width=\hsize]{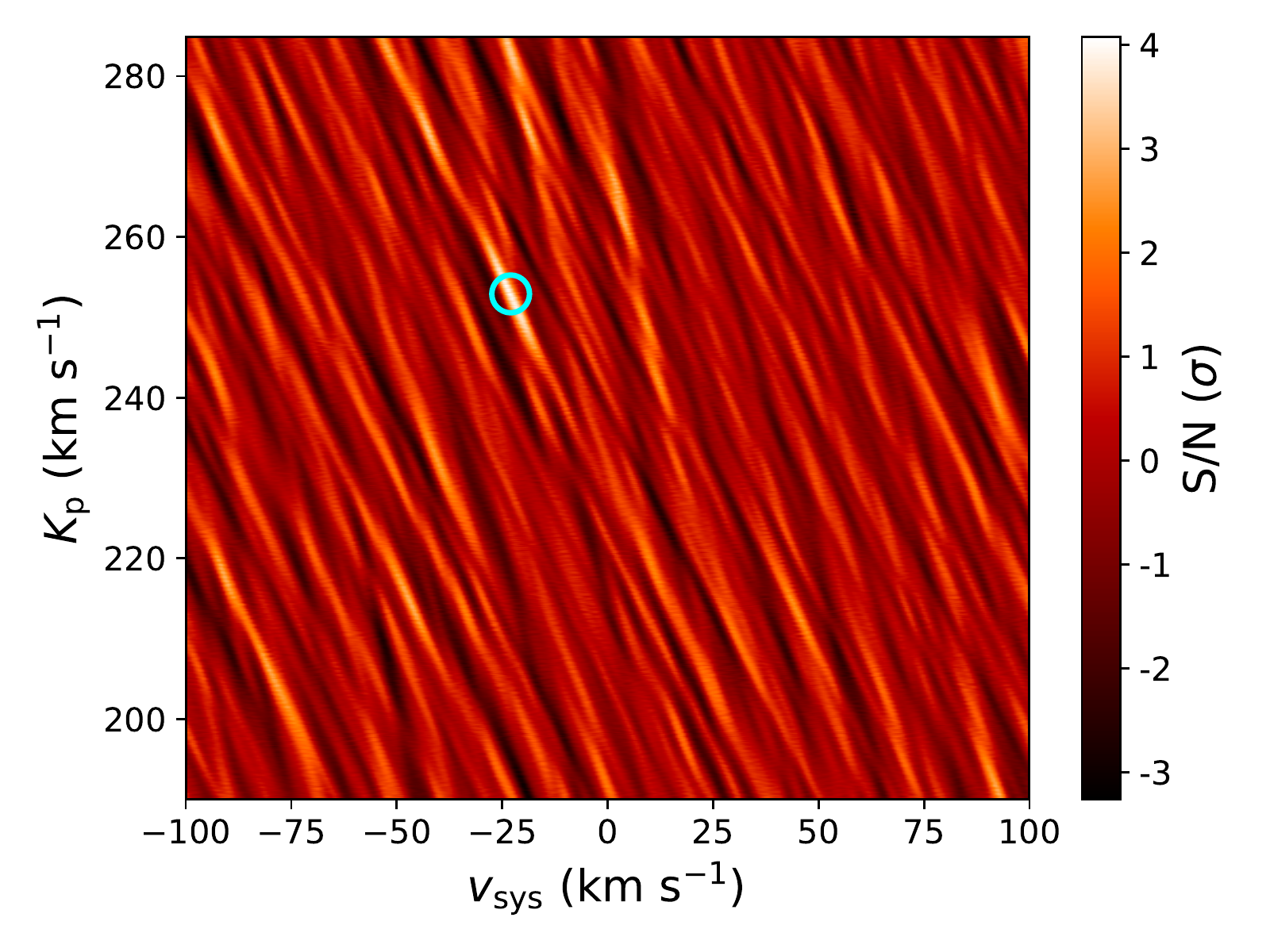}
    \caption{\kp{}-\vsys{} matrix of \snr{} values for the SVD6 data matrix and the ${\lvmr{TiO}=-9.0}$ model. For clarity, the \vsys{}-axis is restricted to $\pm$100 \kms{}. The matrix is plotted on the full \vsys{} range in Figure \ref{fig:snrGridPanel2_full_exomol}. The cyan ring indicates the location of the highest \snr{} value.}
    \label{fig:snrGridSingle}
\end{figure}

    The strongest \snr{} value consistent with our search criteria is that at \vpair{=}{252.9^{+5.0}_{-5.3}}{-23.0^{+4.7}_{-4.6}} for the SVD6 data matrix and the ${\lvmr{TiO}=-9.0}$ model. The corresponding \kp{}-\vsys{} matrix, zoomed to a narrow \vsys{} range for clarity, is shown in Figure \ref{fig:snrGridSingle}. The one-dimensional CCF for ${\kp = 252.9 \ \kms{} }$ is shown in the top panel of Figure \ref{fig:ccf1D}, recomputed over velocities ${\pm 2775\ \kms{} }$ from the planet rest frame. We recalculated the noise as the standard deviation of the CCF across this wider velocity range while excluding the central $\pm$25 \kms{} to derive a more robust ${\snr{}=4.3\sig{}}$ for the peak. The prominence of the central peak is clear over this extensive range of offset velocities.
    
    In the bottom panel of Figure \ref{fig:ccf1D}, we sequentially summed the CCFs of all observations in bins, performed a median subtraction, and normalized to \snr{} values using the standard deviation outside $\pm$25 \kms{}. The first three curves (solid blue, orange, and green) each consist of ten out-of-occultation observations and have peaks with ${\snr \sim \hbox{2--3}\sig{}}$ within 1.0 \kms{} (two pixels) of zero offset velocity. The fourth curve (solid red), consisting of seven out-of-occultation observations, has a ${\snr{}\sim 3\sig{}}$ peak within 1.0 \kms{} of zero offset velocity. The final curve (dashed black) contains only ingress and fully occulted observations, and lacks a similarly distinct, coherent peak close to zero offset velocity. Thus, the presence or absence of the TiO signal seems consistent with the start of occultation.

    \subsection{Welch's $t$-test} \label{sec:ttest}
    
\begin{figure}
    \centering
    \includegraphics[width=\hsize]{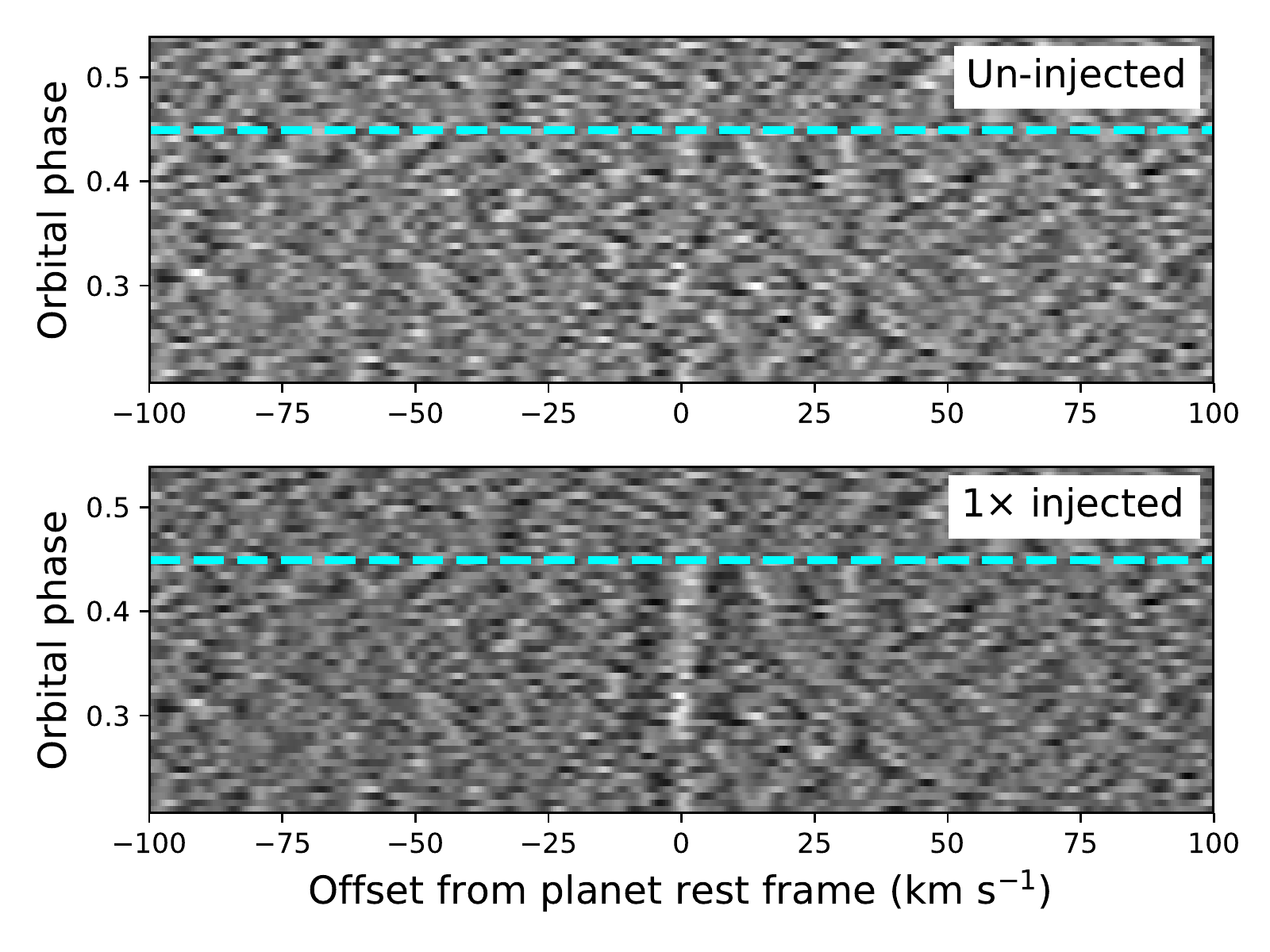}
    \caption{Top panel: Two-dimensional CCF matrix for the ${\lvmr{TiO}=-9.0}$ model and SVD6 data matrix aligned to the exoplanet rest frame for \vpair{=}{252.9}{-23.0}. Each row is the one-dimensional CCF for a given observation. The first observation in occultation is denoted by the dashed cyan line. Orbital phase values 0 and 0.5 correspond to the center of transit and occultation, respectively. Bottom panel: Same for the injection test discussed in Section \ref{sec:injection}. In this case, the velocities \vpair{=}{253.9}{-24.0} were used to shift to the exoplanet rest frame.}
    \label{fig:ccf2D}
\end{figure}

        To further investigate the significance of this emission signal, we performed a two-sided Welch's $t$-test to compare the in- and out-of-trail values of the two-dimensional out-of-occultation CCF matrix for ${\kp{}=252.9\ \kms{}}$ (Figure \ref{fig:ccf2D}, top panel). \citet{Cabot2019} note that the significance given by the Welch's $t$-test is susceptible to overestimation from pixel oversampling. Therefore, we binned the two-dimensional CCF matrix along the velocity axis. Since the resolution element of the HDS data is 1.8 \kms{} and the pixel width is 0.5 \kms{}, we averaged the CCF values in three-pixel-wide bins. We adopted in-trail bounds ${|v|\le 1.5\ \kms{}}$ and out-of-trail bounds ${1.5\ \kms{} < |v| \le 2775\ \kms{}}$.
        
\begin{figure}
    \centering
    \includegraphics[width=\hsize]{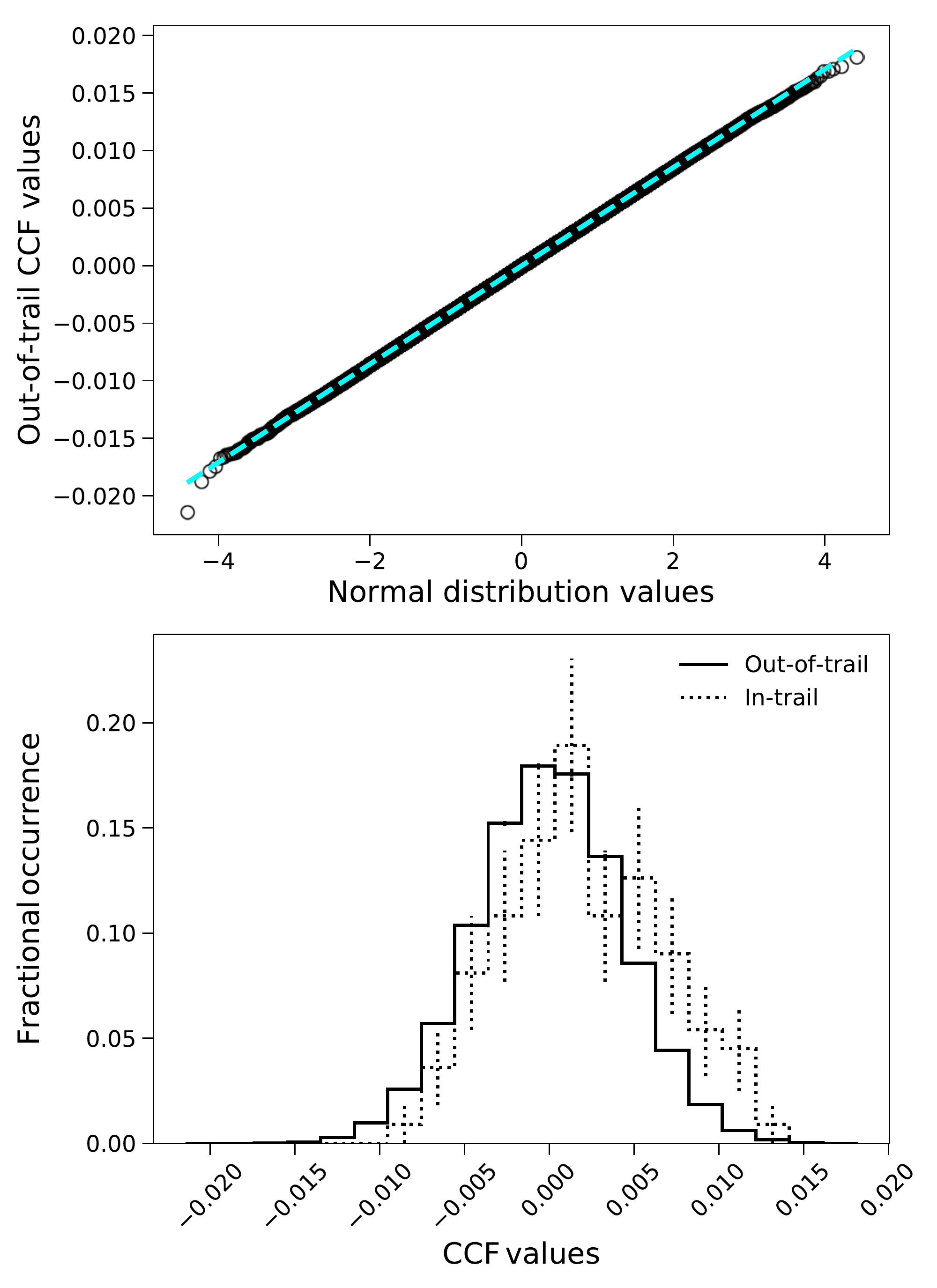}
    \caption{Top panel: Probability plot comparing the binned out-of-trail CCF values to a sampling of the normal distribution (black rings). The dashed cyan line is the calculated linear regression. That the correlation so closely follows a linear trend indicates the out-of-trail values are well described by a Gaussian out to ${\pm4\sig{}}$. Bottom panel: Histogram of the binned in- (dotted line) and out-of-trail (solid line) CCF values. Error bars were calculated as the square root of the individual bin count, normalized by the sample size. A two-sided Welch's $t$-test gives a 4.3\sig{} rejection of the null hypothesis that both populations are draw from the same distribution.}
    \label{fig:stats}
\end{figure}

        First, we verified that the out-of-trail values are Gaussian using the probability plot shown in the top panel of Figure \ref{fig:stats}. Within ${\pm 4 \sig{}}$, the out-of-trail CCF values are tightly correlated to a Gaussian normal distribution sampling. The Welch's $t$-test gives a two-sided $p$-value of ${1.5\times10^{-5}}$, which corresponds to a 4.3\sig{} rejection of the null hypothesis that the in- and out-of-trail CCF values are drawn from the same distribution. This matches the peak \snr{} of the one-dimensional CCF reported in Section \ref{sec:results}.
        
\begin{figure}
    \centering
    \includegraphics[width=\hsize]{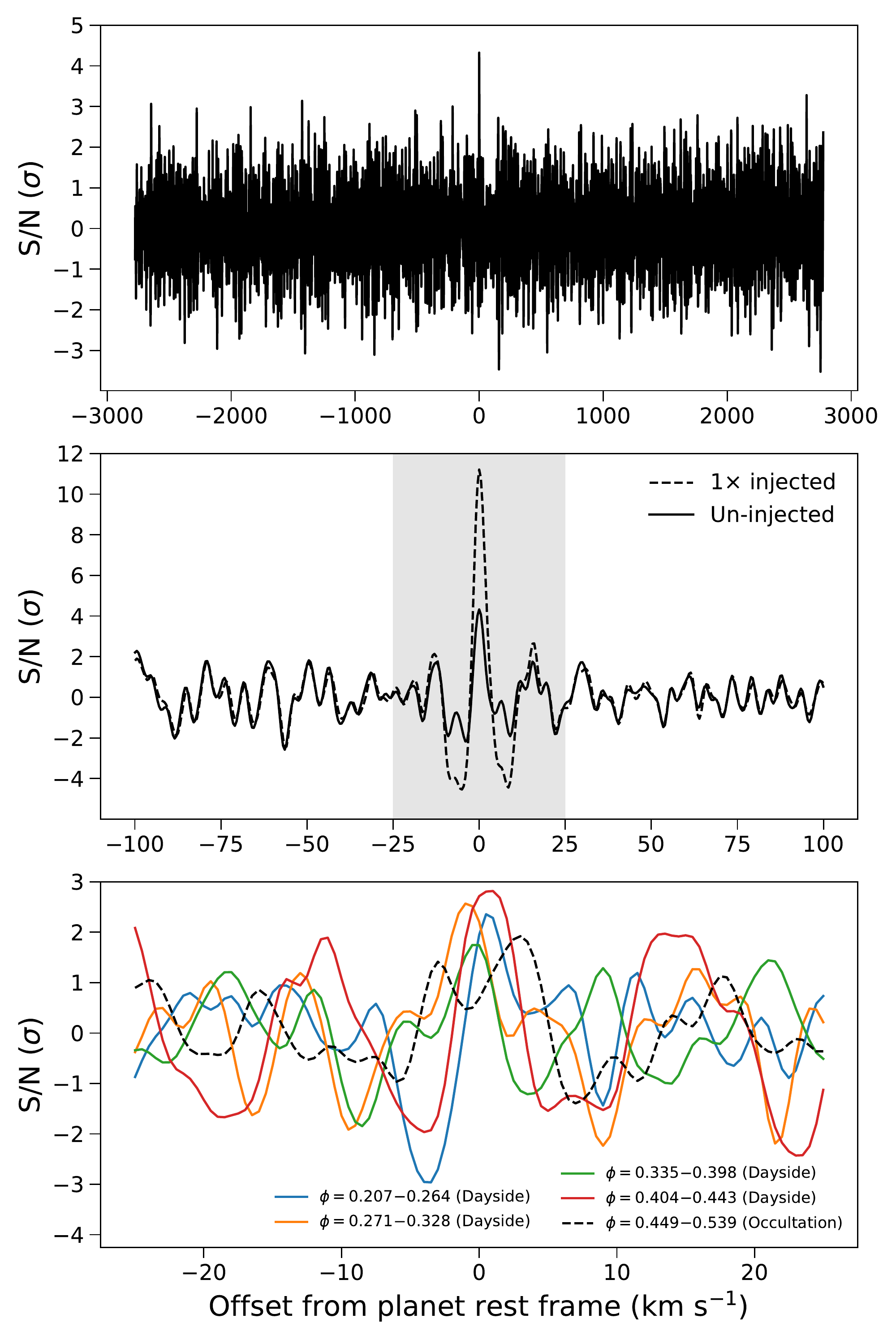}
    \caption{ Top panel: One-dimensional out-of-occultation CCF for the ${\lvmr{TiO} = -9.0}$ model and the SVD6 data matrix aligned to \vpair{=}{252.9}{-23.0}. The CCF values were converted to \snr{} values by dividing by the standard deviation outside ${\pm 25 \ \kms{}}$.
    Middle panel: Same, but on a narrower offset velocity range. The solid line plots the un-injected curve shown in the top panel, while the dashed line denotes the result of the 1$\times$ injection case aligned to \vpair{=}{253.9}{-24.0}. The shaded region denotes the ${\pm 25\  \kms{} }$ range that was excluded from the noise calculation. The peak \snr{} of the injected case is 2.6$\times$ that of the un-injected case.
    Bottom panel: Same as top panel, but instead of combining the out-of-occultation CCFs into a single one-dimensional CCF, we sequentially summed the individual CCFs for various portions of the observing night. Observations 1--37 are completely out of occultation. The CCFs of observations 1--30 are summed in sets of ten and denoted by the solid blue, orange, and green lines. The solid red line shows the combined CCF of observations 31--37. Observations 38--52 contain ingress and full occultation, and the combined CCF for these 15 observations is shown in the dashed black line.
    }
    \label{fig:ccf1D}
\end{figure}

        The bottom panel of Figure \ref{fig:stats} separately plots the in- and out-of-trail CCF values, clearly showing an overall positive shift of the in-trail values relative to the out-of-trail values. We emphasize that the result of the Welch's $t$-test strongly depends on the adopted width of the in-trail region. Our treatment of the data matrices lowers the CCF values immediately surrounding the signal around ${\pm \hbox{2--10}\ \kms{}}$ (Figure \ref{fig:ccf1D}, middle panel), lowering the Welch's $t$-test significance as more of this region is included.
        
    \subsection{Injection-recovery test} \label{sec:injection}
    
        We also performed an injection-recovery test to check if we could successfully recover the planet signal. Using the emission peak velocities \vpair{=}{252.9}{-23.0} determined above, we calculated the relative velocity of WASP-33b at the midpoint of each observation and Doppler shifted the ${\lvmr{TiO} = -9.0}$ model accordingly. We scaled each model emission spectrum to match our data's normalization, accounting for the planet-to-star full-disk ratio and the mid-observation visible illumination fraction of the planet, which we modeled as a Lambertian sphere. After broadening each model planet spectrum to the resolving power of the data and performing an additional boxcar smoothing to simulate the relatively long 600-s integration time, we added our artificial signal to the pre-flagged HDS data. We thereby injected a TiO emission signal at 1$\times$ the expected level for WASP-33b. We subsequently performed the same procedure to reduce stellar, telluric, and systematic noise outlined in Section \ref{sec:data}, removing six SVD components to mimic the treatment that gave our strongest signal. Using the ${\lvmr{TiO} = -9.0}$ template, broadened and filtered as described in Section \ref{sec:petit}, we calculated a \kp{}-\vsys{} matrix as discussed in Section \ref{sec:ccf}.
        
        The location of the strongest peak on this matrix is \vpair{=}{253.9}{-24.0}, matching very closely the injected values. The correspondingly aligned two-dimensional CCF matrix is shown in the bottom panel of Figure \ref{fig:ccf2D}, demonstrating how the injected planet signal increases toward occultation as a larger fraction of the dayside becomes visible along our line of sight. The middle panel of Figure \ref{fig:ccf1D} plots the summed one-dimensional CCF on a restricted range for the injected and un-injected cases, which show minimal difference for velocity offsets outside ${\pm25\ \kms{} }$. Within ${\pm 25\ \kms{}}$, the shape of the central peak and aliasing of the CCF of the injected case matches very closely that of the un-injected case.
        
        As for the un-injected case, we adopted as noise the standard deviation of the one-dimensional CCF over velocities ${\pm2775\ \kms{}}$, excluding the central ${\pm25\ \kms{} }$. We then recover the 1$\times$-injected TiO emission at the 11.2\sig{} level, which is 2.6$\times$ as strong as the un-injected case (4.3\sig{}). Assuming our injected model perfectly represents WASP-33b, we would expect a 1$\times$ injection at the same location as our tentative detection to result in a doubling of the retrieved \snr{}. As our model is not perfect, however, an increase in \snr{} of over 100\% from a 1$\times$ injection-recovery test is to be expected.
        
\section{Discussion and conclusions} \label{sec:discuss}

    Our detection of a tentative TiO signal in the dayside of WASP-33b is intriguing when compared to similar previous studies. Using the new and more accurate \toto{} line list for TiO, we find a potential signal at the 4.3\sig{} level in the HDS spectra. \citet{Nugroho2017} used the older \plez{} TiO line list to recover a stronger 4.8\sig{} TiO signal from the same data. Meanwhile, \citet{Herman2020} analyzed CFHT transit and dayside spectra using the \plezN{} line list, but do not detect TiO. There are four possible scenarios: (1) only our TiO detection with the \toto{} template is valid, (2) only the TiO detection by \citet{Nugroho2017} using the \plez{} line list is valid, (3) both detections are valid, and (4) both detections are false positives. We discuss each below.

    \subsection{Scenario I: Only our detection with \toto{} is valid}
    
        Taken in isolation, there are several aspects of our analysis that make our 4.3\sig{} TiO signal compelling. In addition to the statistical significance of the one-dimensional CCF peak (Figure \ref{fig:ccf1D}, top panel), the Welch's $t$-test presented in Section \ref{sec:ttest} rejects the hypothesis that the in- and out-of-trail CCF samples are drawn from the same distribution. Furthermore, the injection-recovery test discussed in Section \ref{sec:injection} and shown in the middle panel of Figure \ref{fig:ccf1D} lends confidence to our result, in that the shape of the one-dimensional CCF around the signal peak for the 1$\times$-injected case seems to be a scaled version of that for the un-injected case. Additionally, when the aligned two-dimensional CCF is summed in bins (Figure \ref{fig:ccf1D}, bottom panel), the presence or absence of the TiO signal seems consistent with the start of occultation.
    
        Nonetheless, we anticipated our analysis of the same HDS data of WASP-33b using the improved \toto{} line list would yield a stronger TiO detection than the 4.8\sig{} result from \citet{Nugroho2017} using the older \plez{} line list. We initially suspected our weaker detection with a better line list may be partly due to our treatment of the spectral orders. Whereas \citet{Nugroho2017} separately reduced each order and subsequently optimized the number of \sysrem{} iterations performed on an order-by-order basis using injection-recovery tests, we combined all orders into a single spectrum prior to our comparable step of SVD. As a result, our approach has fewer parameters to optimize and may be inherently unable to maximize the \snr{} to a similar degree.
        
        To test whether combining all orders into a single spectrum leads to an inherent decrease in recoverable \snr{}, we performed a data reduction similar to that of \citet{Nugroho2017} that treats each spectral order separately. After the extraction, wavelength drift correction, and initial 5\sig{} clipping steps described in Section \ref{sec:data}, we performed the \sysrem{} detrending algorithm \citep{Tamuz2005} independently on each spectral order. After cross-correlation, the CCFs for each order were added with a uniform \sysrem{} iteration. We chose not to optimize the \sysrem{} iteration for each order based on injection-recovery tests so as to prevent biasing the signal recovery. However, running \sysrem{} on each spectral order separately should, in principle, still facilitate a better reduction, assuming the noise varies for different orders. Similarly, the fact that the \sysrem{} algorithm allows for weighting of individual pixels by their uncertainty during detrending should also enable a more optimized signal recovery. While we readily recover the \citet{Nugroho2017} signal with this methodology when the cross-correlation is performed with their \plez{} models, the results from cross-correlation with the ${\lvmr{TiO}=-9.0}$ \toto{} model are ambiguous and we cannot claim a detection---much less an improvement in signal---using this better-optimized methodology.
        
        Additionally, the \kp{} and \vsys{} values of our TiO signal are offset compared to previous studies. In Figure \ref{fig:velocityError}, we plot our reported velocity values (circular marker), as well as those from \citet{Nugroho2017} (cross marker), \citet{Nugroho2020} (star marker), and other studies of WASP-33b. In addition to a higher value for \kp{}, the value our TiO analysis derives for \vsys{} is blue-shifted compared to values presented in previous works. Particularly intriguing is that the error contours indicate a velocity discrepancy between the TiO results of this work and \citet{Nugroho2017}, but may indicate agreement between our TiO result and the Fe {\sc I} detection of \citet{Nugroho2020}. We emphasize that all three analyses used the same HDS data of WASP-33b. Also puzzling is the fact that we derive a ${\vsys{} \sim -3\ \mathrm{to}\ -5\ \kms{}}$ based solely on cross-correlating a rotationally broadened stellar model and the HDS data, which is comparable to previous studies.
    
    \subsection{Scenario II: Only the \cite{Nugroho2017} detection with \plez{} is valid}
    
\begin{figure}
    \centering
    \includegraphics[width=\hsize]{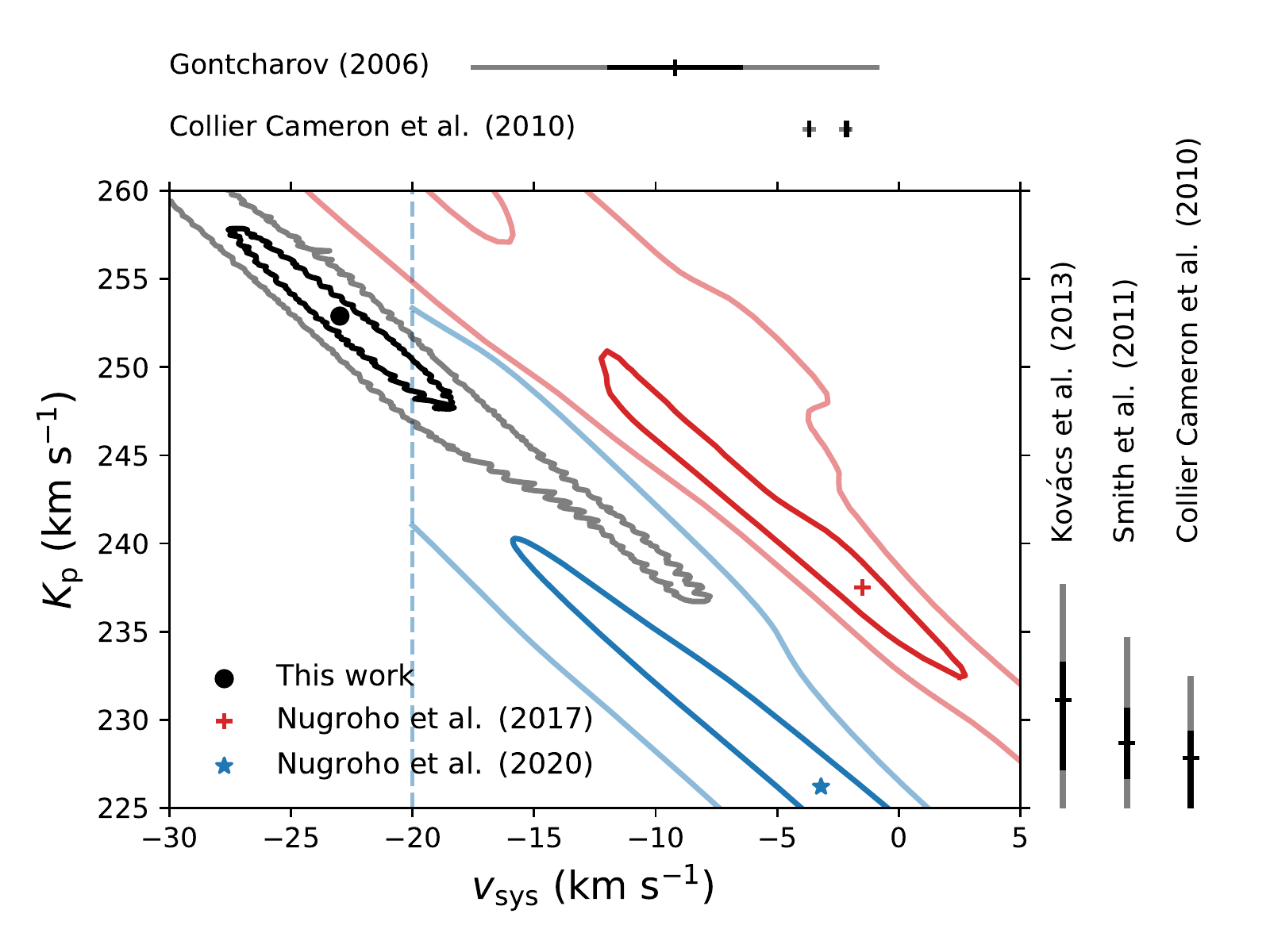}
    \caption{Comparison of various reported values of \kp{} and \vsys{} for the WASP-33 system. Our result is indicated by the circular marker, along with 1\sig{} and 3\sig{} contours in black and gray, respectively. The corresponding results from Figure 14 in \citet{Nugroho2017} for TiO are also plotted, with the location of their \snr{} peak denoted by the cross marker and the 1\sig{} and 3\sig{} contours shown in dark and light red. The \snr{} peak from the recent \citet{Nugroho2020} Fe {\sc I} detection is indicated by the star marker, with the 1\sig{} and 3\sig{} contours from the \snr{} matrix shown as solid dark blue and solid light blue lines. The ${\vsys{}=-20\ \kms{}}$ lower limit in their analysis caused by stellar pulsations is denoted by the dashed blue line. Previously reported values for \vsys{} \citep{Gontcharov2006,CollierCameron2010} and \kp{} \citep{Kovacs2013,Smith2011,CollierCameron2010} are shown above and to the right of the plot, with the strokes indicating the reported values, and the extended black and gray bars denoting the 1\sig{} and 3\sig{} errors, respectively. The negative errors for the \kp{} values are clipped for clarity.}
    \label{fig:velocityError}
\end{figure}

        In light of these issues, another possible scenario is that the \citet{Nugroho2017} result is valid. While the corresponding values of \kp{} and \vsys{} that \citet{Nugroho2017} report are indeed in better agreement with previous works (Figure \ref{fig:velocityError}), we were unable to reproduce the same unambiguous detection of TiO emission in WASP-33b using our data treatment with their \plez{} model templates. For reference, we include the corresponding \kp{}-\vsys{} matrices in Figures \ref{fig:snrGridPanel1_plez9}--\ref{fig:snrGridPanel2_plez9} and Figures \ref{fig:snrGridPanel1_plez8}--\ref{fig:snrGridPanel2_plez8} for the ${\lvmr{TiO}=-9.0,\ -8.0}$ cases, which are the fiducial volume mixing ratios from this work and \citet{Nugroho2017}, respectively.
    
        To estimate the recoverability of the TiO signal in WASP-33b with our methodology and the \citet{Nugroho2017} models based on the \plez{} line list, we performed the same \toto{} injection described in Section \ref{sec:injection} except we cross-correlated with a \plez{} TiO emission template from \citet{Nugroho2017}. This way, we could compare the signal lost due to differences in the models. For the recovery, we used the ${\lvmr{TiO}=-9.0}$ contrast model presented in \citet{Nugroho2017}, broadened and filtered in the same manner as our \toto{} models to match our treatment of the dataset.
    
        The resulting global peak on the SVD6 \kp{}-\vsys{} matrix for this \plez{} recovery of injected \toto{} signal is \vpair{=}{255.4}{-24.0}, which is very close to the injected values. This demonstrates that the aforementioned \vsys{} discrepancy between this work and \citet{Nugroho2017} is not caused by differences in the model templates. Indeed, cross-correlating these ${\lvmr{TiO}=-9.0}$ models reveals a relative velocity offset of only 2.0 \kms{}, while cross-correlating the ${\lvmr{TiO}=-9.0}$ \toto{} and ${\lvmr{TiO}=-8.0}$ \plez{} models indicates a relative velocity offset of just 1.5 \kms{}. As before, we calculated the noise on the correspondingly aligned one-dimensional CCF between $\pm 2775 \ \kms{}$, excluding the inner ${\pm 25\ \kms{}}$, to get a 5.1\sig{} detection with the \plez{} model from \citet{Nugroho2017}. This represents a 54\% loss compared to the 11.2\sig{} signal recovered with our \toto{} model. Scaling our 4.3\sig{} detection using the \toto{} models and un-injected data, we estimate the TiO emission present in WASP-33b to be recoverable only at the ${\lesssim2\sig{}}$ level using the \plez{} spectral models from \citet{Nugroho2017} with our methodology.
        
    \subsection{Scenario III: Both detections are valid}
    
        A third possible, but unlikely, scenario is that our TiO emission signal found with the \toto{} spectral templates and the signature reported by \citet{Nugroho2017} using \plez{} models are both valid. In principle, the comparative completeness of the line lists may influence the retrieved signal. An increased quantity of weak TiO lines in our \toto{} model template may decrease the effective contrast of the TiO emission, thereby degrading the \snr{}. Our fixation of the pressure-temperature profile may also be a contributing factor. Differences in line strengths, line positions, and overall completeness between the \toto{} and \plez{} line lists may lead to substantially different line profiles even for similar atmospheric models (pressure-temperature profile, mean molecular weight profile, TiO volume mixing ratio, etc.), potentially leading to a weaker detection with the \toto{} models. Indeed, the normalized cross-correlation of the filtered $\lvmr{TiO}=-9.0$ \plez{} and \toto{} models peaks at a value of only 0.28, indicating substantial differences in the spectral models. Similar points are raised by \citet{Gandhi2020} when interpreting a discrepancy in detecting methane using two different line lists. 
        
        We attempted to recover our \toto{} TiO signal using models based on different pressure-temperature profiles to determine whether the signal strength substantially changes for different atmospheric structures. As limiting cases, we adopted fully inverted and non-inverted pressure-temperature profiles that span the same pressure (10$^2$--10$^{-5}$ bar) and temperature (2700--3700 K) ranges as the \citet{Haynes2015} profile, but vary monotonically with a constant lapse rate. These are the same fully and non-inverted profiles used by \citet{Nugroho2017}. We followed the same procedure presented in Section \ref{sec:methods} to create model spectra with \ptr{} for the two alternate atmospheric profiles, and subsequently cross-correlated these models with the processed data. In both the fully and non-inverted cases, varying the total TiO abundance does not lead to significant differences in the final filtered models compared to that for the fiducial abundance of $\lvmr{TiO}=-9.0$, so we investigated only with the latter abundance. The fully inverted and non-inverted models recover very similar TiO emission signals, with the former giving a 4.3\sig{} correlation and the latter giving a $-4.4$\sig{} anti-correlation at \kp{} and \vsys{} values consistent with the 4.3\sig{} peak recovered using the \citet{Haynes2015} model. A more exhaustive investigation might determine an optimal inverted pressure-temperature profile that enables the recovery of a stronger TiO emission signal for models based on the \toto{} line list, but such a study is beyond the scope of this work.
    
        Nonetheless, the various points previously raised in this section pose serious challenges to this interpretation that both detections are valid. A degradation in contrast by additional weak lines and our fixation of the pressure-temperature profile would not explain our inability to recover the \citet{Nugroho2017} detection using their \plez{} models. Furthermore, we were unable to find an explanation for the offset in reported \vsys{}.
    
    \subsection{Scenario IV: Both detections are false positives}
    
        The fourth possible scenario is that both TiO emission signals are in fact false positives. \citet{Herman2020} do not find evidence for TiO in either dayside or transmission spectra of WASP-33b. This interpretation has the advantage of explaining the noted discrepancies between this work and \citet{Nugroho2017}. Such a scenario would raise questions regarding the current statistical techniques used to evaluate detection significance in high-resolution spectroscopy, as well as highlight the importance of reanalyzing detections using a diverse set of reduction methodologies and spectral models. In the end, further high-resolution spectra may be required to resolve the ambiguity surrounding the presence of TiO in WASP-33b.


\begin{acknowledgements}
DBS, PM, and IAGS acknowledge support from the European Research Council under the European Union’s Horizon 2020 research and innovation program under grant agreement No. 694513.

SKN would like to acknowledge support from UK Science Technology and Facility Council grant ST/P000312/1.

PM acknowledges support from the European Research Council under the European Union’s Horizon 2020 research and innovation program under grant agreement No. 832428.

NPG gratefully acknowledges support from Science Foundation Ireland and the Royal Society in the form of a University Research Fellowship.
\end{acknowledgements}


\bibliographystyle{aa}
\bibliography{tio_wasp33b.bib}


\begin{appendix}

    \section{\kp{}-\vsys{} matrices}
    
    \begin{figure*}[h!]
        \centering
        \includegraphics[width=\hsize]{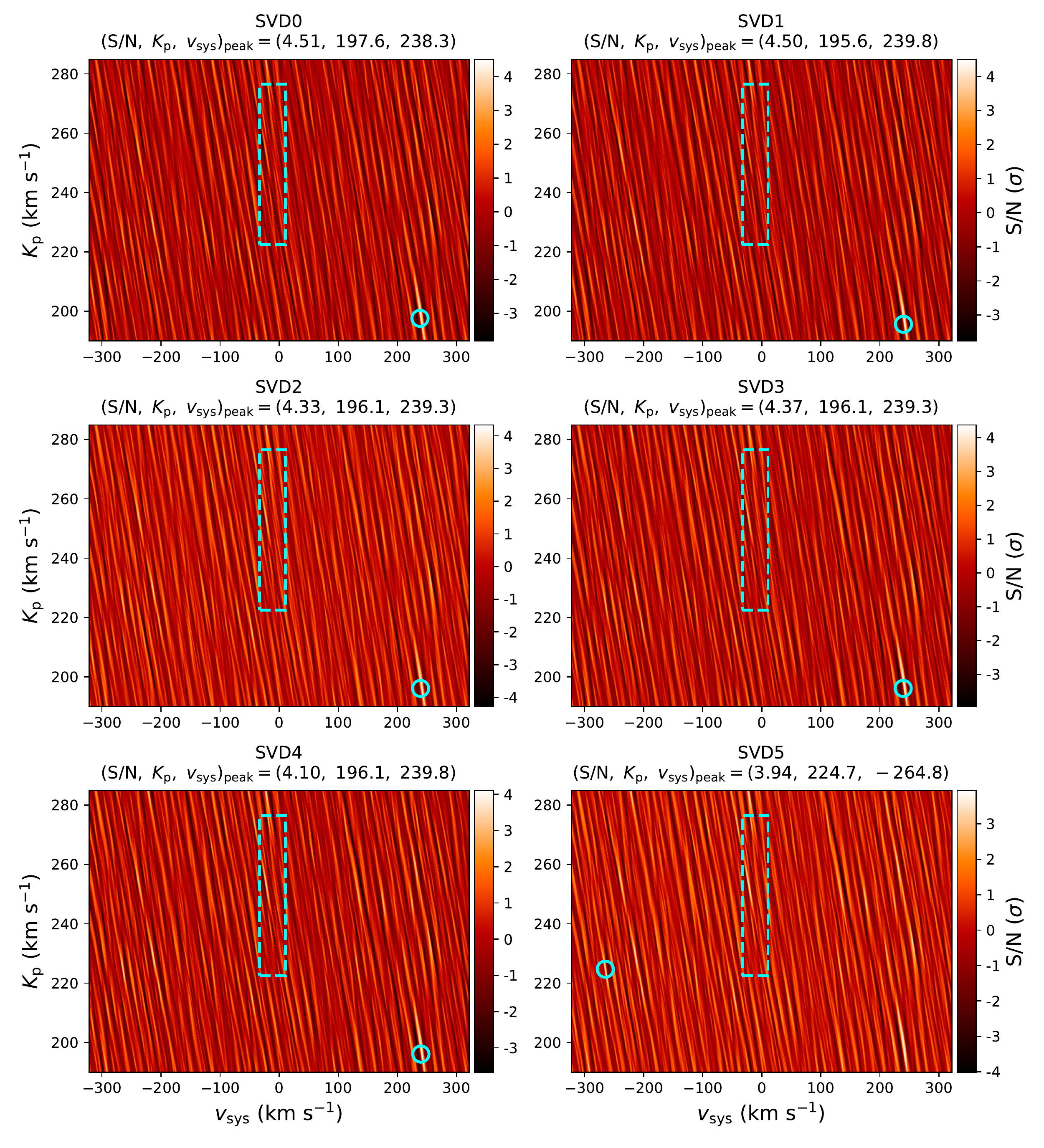}
        \caption{\kp{}-\vsys{} matrices of \snr{} values for the ${\lvmr{TiO}=-9.0}$ \toto{} model, as described in Section \ref{sec:ccf}. Each panel corresponds to the matrix for a different SVD iteration (0--5), indicated above each panel. The strongest peak is marked on each matrix with a cyan ring and noted above each panel with the corresponding values of \kp{} and \vsys. The dashed cyan box indicates the region where peaks must be found to be considered "significant," as described in Section \ref{sec:results}. }
        \label{fig:snrGridPanel1_full_exomol}
    \end{figure*}
    
    \newpage
    
    \begin{figure*}
        \centering
        \includegraphics[width=\hsize]{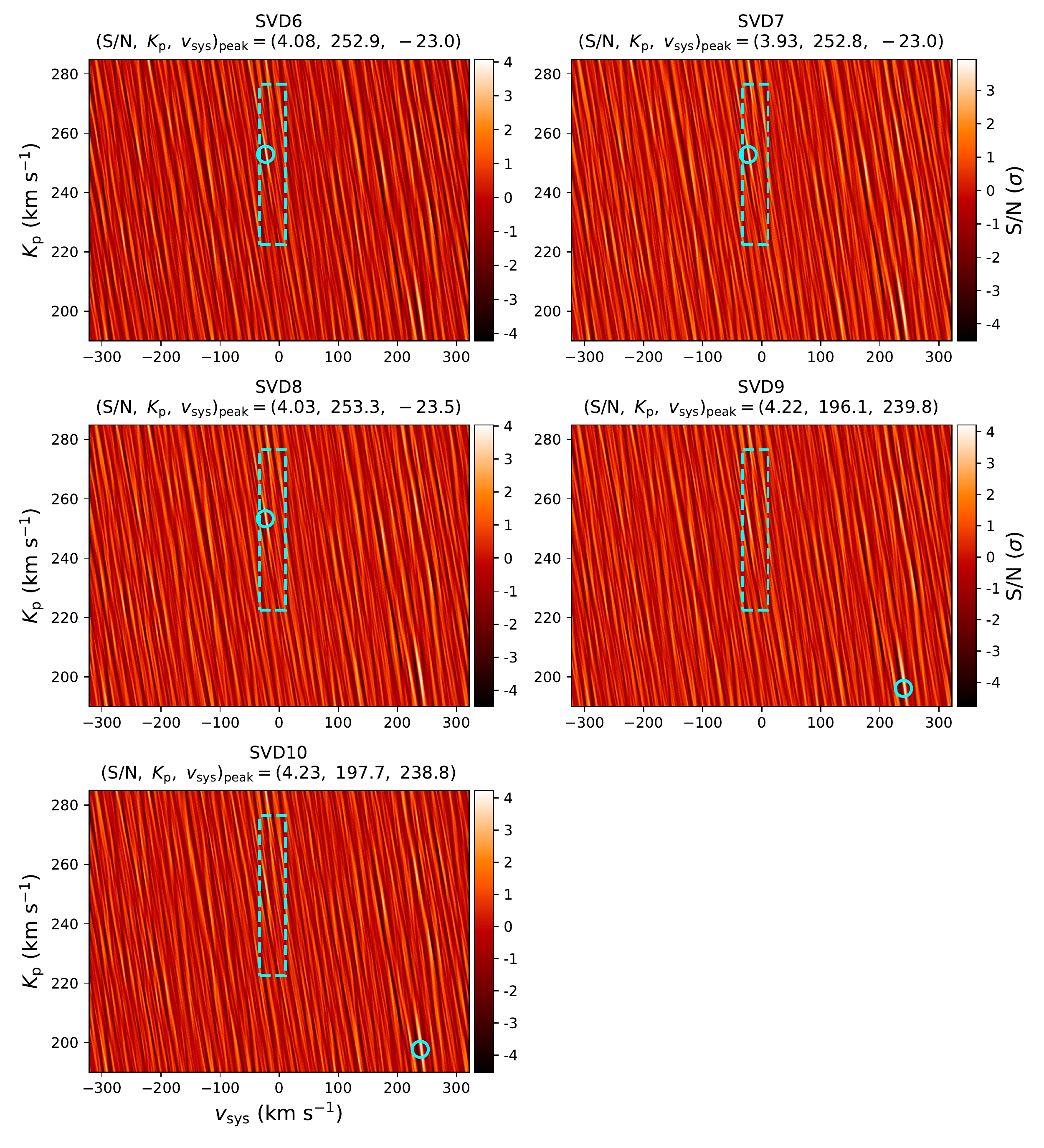}
        \caption{Same as Figure \ref{fig:snrGridPanel1_full_exomol}, but for SVD iterations 6--10.}
        \label{fig:snrGridPanel2_full_exomol}
    \end{figure*}
    
    \newpage
    
    \begin{figure*}
        \centering
        \includegraphics[width=\hsize]{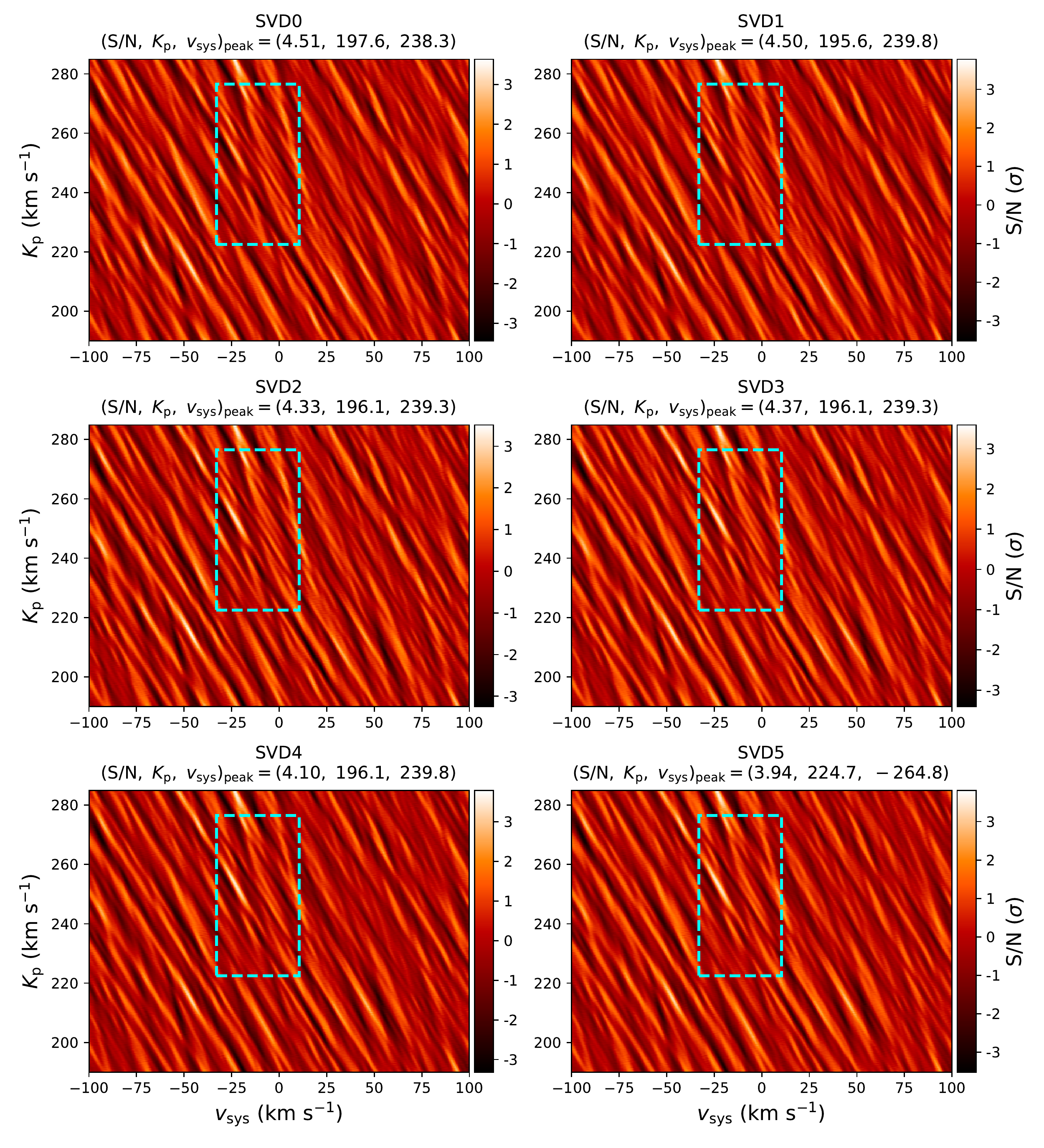}
        \caption{Same as Figure \ref{fig:snrGridPanel1_full_exomol}, but with the \vsys{}-axis restricted to values within $\pm$100 \kms{} for clarity. While the \snr{} peak printed above each panel refers to that for the full \vsys{} range, the colorbar scaling is based on the \snr{} values on this restricted \vsys{} range.}
        \label{fig:snrGridPanel1_zoom_exomol}
    \end{figure*}
    
    \newpage
    
    \begin{figure*}
        \centering
        \includegraphics[width=\hsize]{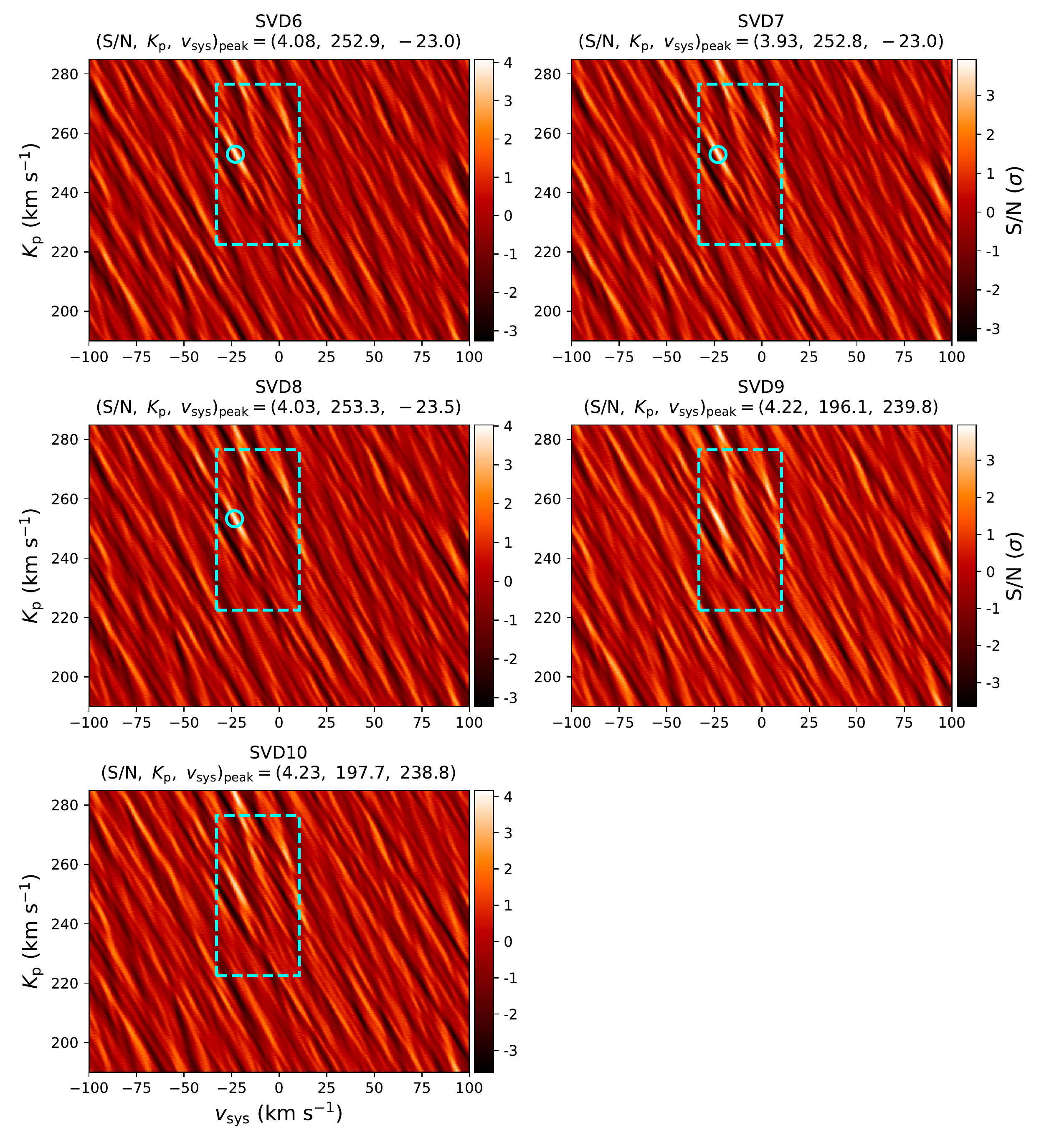}
        \caption{Same as Figure \ref{fig:snrGridPanel1_zoom_exomol}, but for SVD iterations 6--10.}
        \label{fig:snrGridPanel2_zoom_exomol}
    \end{figure*}
    
    \newpage
    
    \begin{figure*}
        \centering
        \includegraphics[width=\hsize]{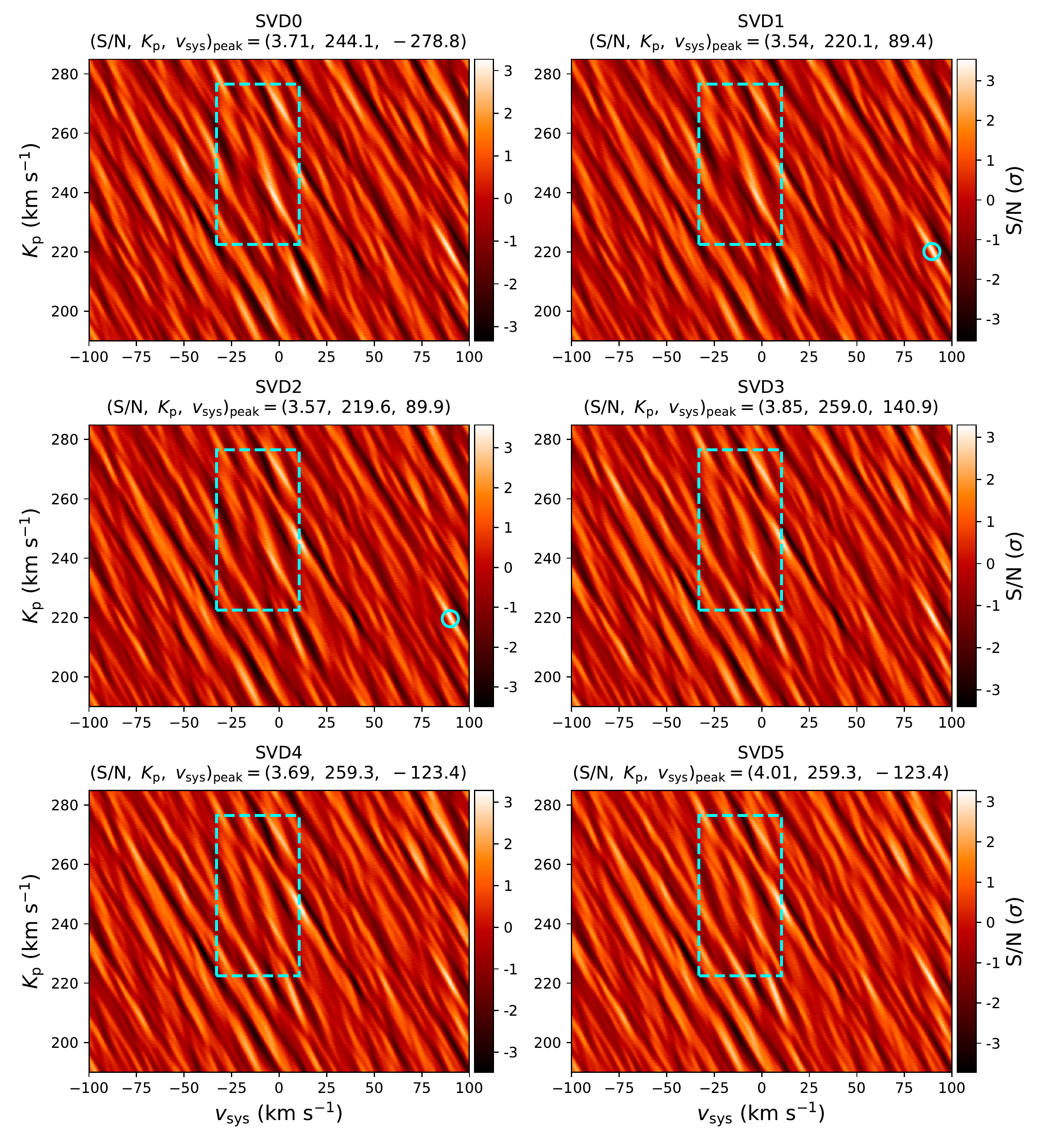}
        \caption{\kp{}-\vsys{} matrices of \snr{} values for the ${\lvmr{TiO}=-9.0}$ \plez{} model from \citet{Nugroho2017}, with the \vsys{}-axis restricted to values within $\pm$100 \kms{} for clarity. Each panel corresponds to the matrix for a different SVD iteration (0--5), indicated above each panel. The strongest peak on the full \vsys{} range is noted above each panel with the corresponding values of \kp{} and \vsys{}. Where applicable, this peak is also marked on each matrix with a cyan ring. The dashed cyan box indicates the region where peaks must be found to be considered significant, as described in Section \ref{sec:results}. The colorbar binning is based on the local \snr{} minimum and maximum on this restricted \vsys{} range.}
        \label{fig:snrGridPanel1_plez9}
    \end{figure*}
    
    \newpage
    
    \begin{figure*}
        \centering
        \includegraphics[width=\hsize]{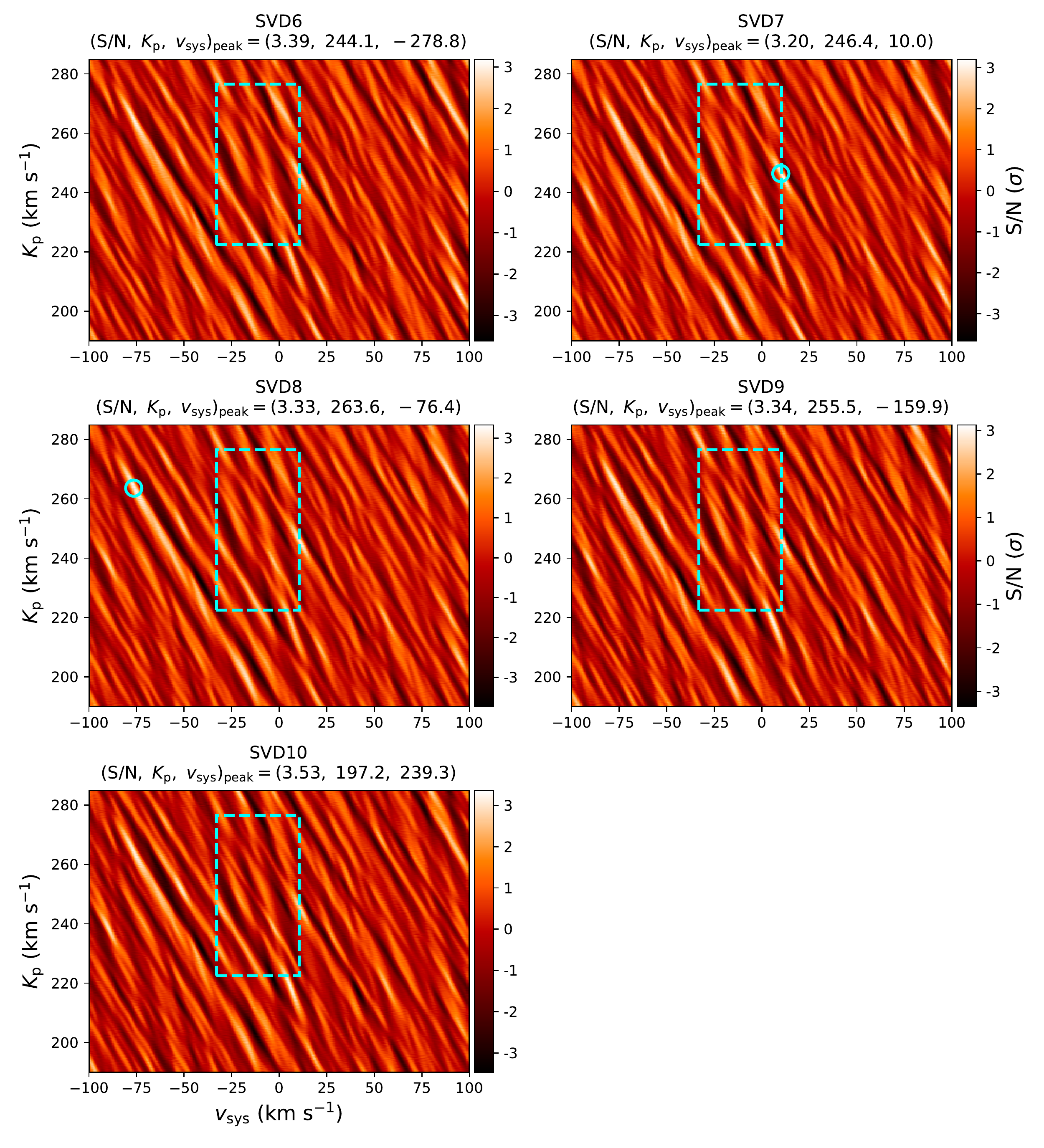}
        \caption{Same as Figure \ref{fig:snrGridPanel1_plez9}, but for SVD iterations 6--10.}
        \label{fig:snrGridPanel2_plez9}
    \end{figure*}
    
    \newpage
    
    \begin{figure*}
        \centering
        \includegraphics[width=\hsize]{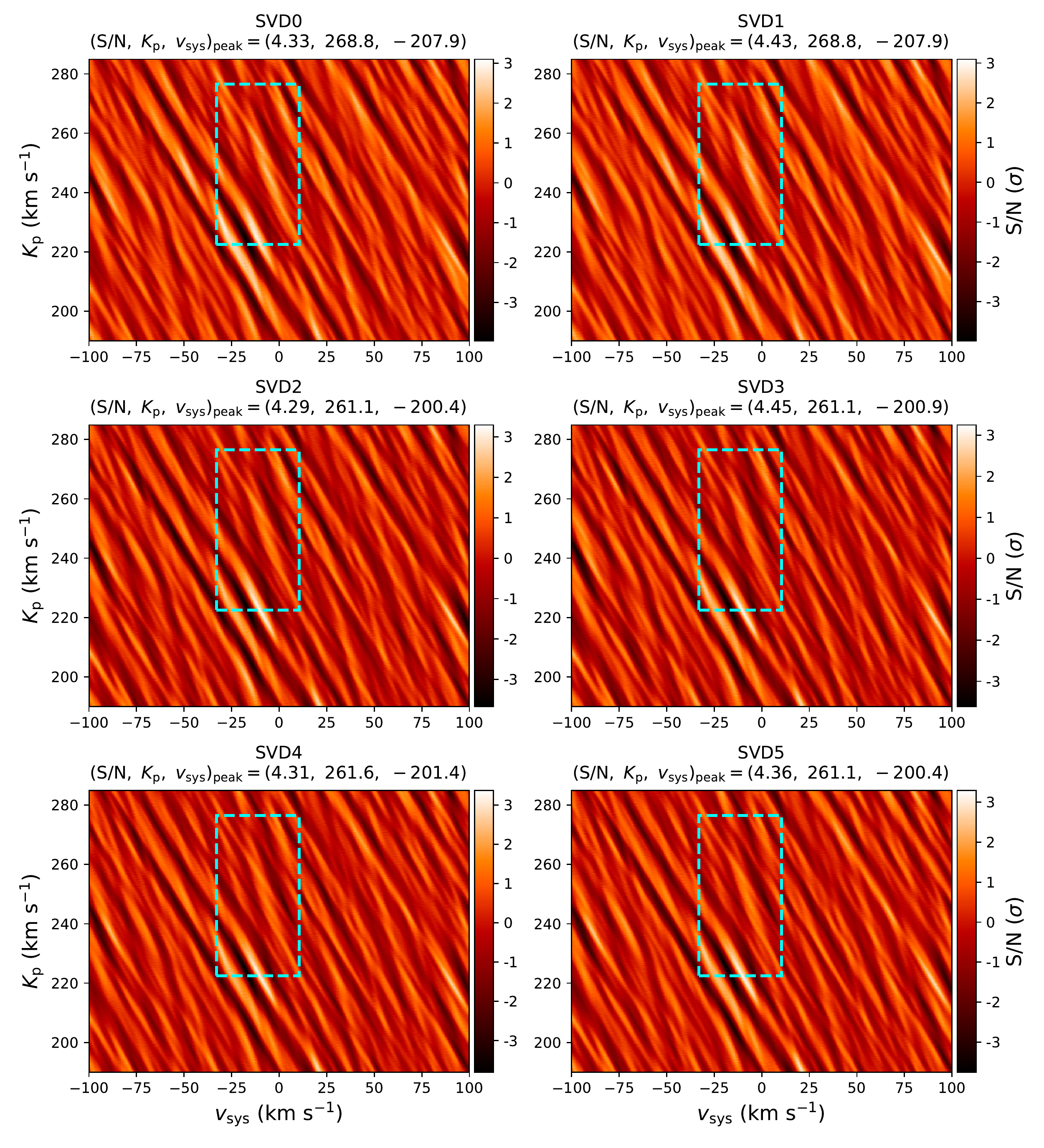}
        \caption{\kp{}-\vsys{} matrices of \snr{} values for the ${\lvmr{TiO}=-8.0}$ \plez{} model from \citet{Nugroho2017}, with the \vsys{}-axis restricted to values within $\pm$100 \kms{} for clarity. Each panel corresponds to the matrix for a different SVD iteration (0--5), indicated above each panel. The strongest peak on the full \vsys{} range is noted above each panel with the corresponding values of \kp{} and \vsys{}. Where applicable, this peak is also marked on each matrix with a cyan ring. The dashed cyan box indicates the region where peaks must be found to be considered significant, as described in Section \ref{sec:results}. The colorbar binning is based on the local \snr{} minimum and maximum on this restricted \vsys{} range.}
        \label{fig:snrGridPanel1_plez8}
    \end{figure*}
    
    \newpage
    
    \begin{figure*}
        \centering
        \includegraphics[width=\hsize]{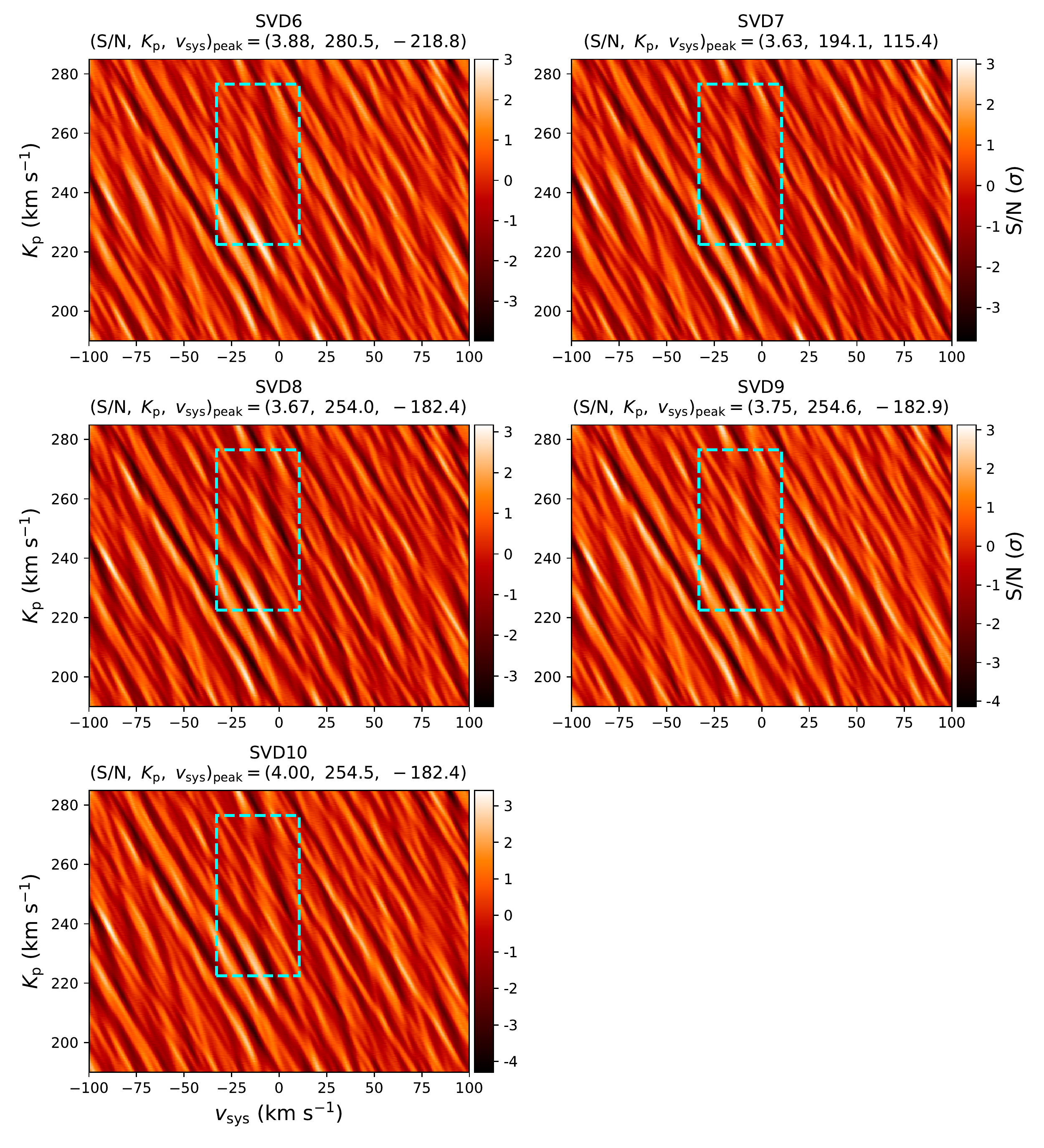}
        \caption{Same as Figure \ref{fig:snrGridPanel1_plez8}, but for SVD iterations 6--10.}
        \label{fig:snrGridPanel2_plez8}
    \end{figure*}

\end{appendix}

\end{document}